\documentclass[pra,twocolumn,superscriptaddress,showpacs]{revtex4-2}

\usepackage{amsfonts}
\usepackage{braket}
\usepackage{units}
\usepackage{mathrsfs}

\pdfoutput=1
\usepackage{amsmath}
\usepackage{bm}
\usepackage{amssymb}
\usepackage{graphicx}
\usepackage{color}
\usepackage{simplewick}
\usepackage[version=3]{mhchem}
\usepackage{wasysym}
\usepackage{physics}

\DeclareMathOperator{\re}{Re}

\begin{document}

\title{Dicke transition in open many-body systems determined by fluctuation effects}

\newcommand{\colt}{Institute for Theoretical Physics,
University of Cologne, 50937 Cologne, Germany}

\newcommand{\bonnpi}{Physikalisches Institut, University of Bonn, Nussallee 12, 53115 Bonn, Germany}

\author{Alla V. Bezvershenko}
\affiliation{\colt}
\author{Catalin-Mihai Halati}
\affiliation{\bonnpi}
\author{Ameneh Sheikhan}
\affiliation{\bonnpi}
\author{Corinna Kollath}
\affiliation{\bonnpi}
\author{Achim Rosch}
\affiliation{\colt}




\maketitle



\textbf{In recent years, one important experimental achievement was the strong coupling of quantum matter and quantum light \cite{FornDiazSolano2019}. Realizations reach from ultracold atomic gases in  high-finesse optical resonators  \cite{RitschEsslinger2013} to electronic systems coupled to THz cavities \cite{FornDiazSolano2019}. 
  The dissipative nature of the quantum light field and the global coupling to the quantum matter leads to many exciting phenomena such as the
   occurrence of dissipative quantum phase transition to self-organized exotic phases \cite{RitschEsslinger2013, FornDiazSolano2019}. 
 Previously, often mean-field approaches were applied which characterize the emergence of self-organized phases as a zero-temperature transition for the particles, a ground-state Dicke transition \cite{RitschEsslinger2013}. Here we develop a new approach which combines a mean-field approach with a perturbative treatment of fluctuations beyond mean-field, which becomes exact in the thermodynamic limit. We argue that these fluctuations are crucial in order to determine the mixed state (finite temperature) character of the transition and to unravel universal properties of the 
 self-organized states. We validate our results by comparing to time-dependent matrix-product-state calculations.
 }

The interfacing of quantum light and quantum matter is an important building block for quantum technological devices. Quantum light is advantageous in order to transport quantum information, whereas quantum matter allows one to control and perform quantum operations. Therefore, the achievement and control of strong coupling of quantum matter and quantum light became a very active field of research both in atom based \cite{BaumannEsslinger2010, KlinderHemmerich2015,RitschEsslinger2013, RouxBrantut2020} or solid state based systems \cite{FornDiazSolano2019}. 
The efficient coupling of light and matter allows to realize experimentally the so-called Dicke transition, a quantum phase transition to a self-organized superradiant state \cite{Dicke1954,RitschEsslinger2013,DamanetKeeling2019}. Originally this phenomenon was discussed in a simplified model of two-level systems in a cavity \cite{Dicke1954}, but
finds now application in a wide range of setups where interacting many-body systems are coupled to the modes of a cavity \cite{RitschEsslinger2013,KlinderHemmerich2015b, LandigEsslinger2016}.

Due to the complexity of the models required to describe the hybrid system, theoretical treatments of coupled atomic cavity systems often resort to  a mean field decoupling of the cavity field and the particles assuming an effective ground state for the particles \cite{RitschEsslinger2013, MaschlerRitsch2008
}. Only few efforts go beyond these mean-field studies including the light matter coupling, e.g.~for non-interacting two-level atoms \cite{
KirtonDallaTorre2019} (and references therein), finite size systems  \cite{VukicsRitsch2007, 
ZhangZhou2008, KramerRitsch2014, SandnerRitsch2015, OstermannRitsch2020, HalatiKollath2020,GammelmarkMolmer2012, WallRey2016} or closed systems \cite{PiazzaZwerger2013,SchulerRabl2020}.

The standard mean-field decoupling \cite{RitschEsslinger2013, MaschlerRitsch2008
} of the cavity and the matter part has a crucial problem: whereas the solution for the cavity field is well defined within the mean-field approach, the solution for the steady state of the quantum matter is not. Typically, a pure state, the ground state of the mean-field Hamiltonian, had been chosen for the matter component, a somewhat arbitrary choice.
Here we overcome this long-standing question of the arbitrariness of the mean-field approach by taking fluctuations induced by the light-matter coupling perturbatively into account. The developed approach is very generally applicable and becomes exact in the thermodynamic limit. We demonstrate its validity at the example of a bosonic quantum gas coupled to an optical cavity comparing it to quasi-exact matrix product state (MPS) calculations \cite{HalatiKollath2020, HalatiKollath2020b}.

  In order to describe 
  interacting particles coupled globally to a dissipative light field, we use  the Liouville equation \cite{RitschEsslinger2013, MaschlerRitsch2008}
\begin{align}
\hat{\mathcal{L}}\rho=-\frac{i}{\hbar}[H,\rho]+\Gamma\left(a \rho a^{\dagger} -\frac{1}{2}\{a^{\dagger}a, \rho\}\right),
\label{eq:lind}
\end{align}
where $a^{(\dagger)}$ is the bosonic annihilation (creation) operator for the light mode and the Lindblad operator $a$ gives the loss from the light mode with strength $\Gamma$. 

The Hamiltonian is of the form, $H=H^0_\text{c}+H^0_\text{b}+H_{\text{bc}}$. Here $H^0_c$ contains only operators of the light field, $H^0_b$ is an interacting many-body Hamiltonian for the particles and,  $H_{\text{bc}} =-\frac{\hbar g}{\sqrt{L}}(a +a^{\dagger})O$, couples the light field to an (extensive) operator $O$ acting on the matter fields only. The prefactor $1/\sqrt{L}$, where $L$ is the size of the system, is necessary to obtain a meaningful thermodynamics limit $L\to \infty$.

 We split the matter-cavity coupling into a mean-field contribution, $H_{bc}^{MF}$ and fluctuations $\delta H_{bc}$, $H_{bc}=H_{bc}^{MF}+\delta H_{bc}+\text{const.}$, with
\begin{align}
H_{bc}^{MF}&=-\hbar g \lambda O- \hbar g \sqrt{L} ( a+a^{\dagger}) \Delta, \label{eq:MF}
\end{align}
where 
\begin{align}
\lambda=&\frac{\langle a + a^\dagger\rangle_c}{\sqrt{L}}, ~\Delta=\frac{\langle O \rangle_b}{L}, \label{eq:MF2}
\end{align}
which need to be computed self-consistently.
 The cavity mode obtains in the superradiant phase an expectation value $\langle a+a^{\dagger} \rangle_c \sim \sqrt{L}$ leading to a mean-field contribution of $O(1)$.
While the contributions of the fluctuations, $\delta H_{bc}$, are small, $O(1/\sqrt{L})$, we show that they become important in the long-time limit.

The mean-field master equation is given  by 
\begin{align}
\hat{\mathcal{L}_0}\rho&=-\frac{\textit{i}}{\hbar}[H_b(\lambda)+H_c(\Delta),\rho]+\Gamma\left(a \rho a^{\dagger} -\frac{1}{2}\{a^{\dagger}a, \rho\}\right) \nonumber\\
\end{align}
with $H_b(\lambda)=H_b^{(0)}-\hbar g \lambda O$ and  $H_c(\Delta)=H_c^{(0)}- \hbar g \sqrt{L} ( a+a^{\dagger}) \Delta$ determined by the self-consistency condition, Eq.~\eqref{eq:MF2}. The corresponding decoherence-free subspace is spanned by states satisfying ~$\hat{\mathcal{L}_0}\rho_0=\lambda_0 \rho_0$, with $\re\lambda_0=0$. We can factorize these states as $\rho_0=\rho_0^b\cdot\rho_0^c$. The steady-state density matrix of the cavity is $\rho_0^c \propto |\alpha \rangle\langle \alpha |$, with the coherent state $|\alpha \rangle \sim e^{\alpha a^\dagger}|0\rangle$ for
 \begin{equation}
  \label{eq:MF_cavity}
 \frac{\alpha}{\sqrt{L}}=\frac{\langle a\rangle_c}{\sqrt{L}} =\frac{  g \Delta}{\delta-i \Gamma/2}.
 \end{equation}
 In contrast, all combinations of eigenstates $\Ket{m}$ of $H_b(\lambda)$, i.e. $\Ket{n}\Bra{m}$, are non-decaying eigenstates of $\hat{\mathcal{L}_0}$ 
 . Thus a general state in the decoherence free subspace is
  \begin{align}
   &\rho(t)= |\alpha (\Delta) \rangle\langle \alpha (\Delta)|\cdot\rho^b(\lambda),~\text{with}~\rho^b=\sum_{n,m} c_{nm}\Ket{n}\Bra{m}.
   \end{align}
Note that the expectation values in the definitions of $\lambda$ and $\Delta$ are taken with respect to $\rho_0$ and are thus determined by the unknown parameters $c_{n,m}$.

Therefore the density matrix of the particle system is {\em not} uniquely fixed by the master equation. 
The conventional approach, used in many studies \cite{RitschEsslinger2013}
, is to consider simply the ground state of $H_b$. This arbitrariness can be resolved by taking fluctuations perturbatively into account using the perturbation \footnote{Similar to the many body adiabatic elimination techniques described in Ref.~\cite{Garcia-RipollCirac2009,ReiterSorensen2012,PolettiKollath2012}.}
\begin{align}
\hat{\mathcal{L}_1}\rho&=-\frac{\textit{i}}{\hbar}[\delta H_{bc},\rho].
\end{align}
This is justified as $\hat{\mathcal{L}_1}$ scales with $1/\sqrt{L}$ and becomes exact in the thermodynamic limit. The perturbative approach then determines the time evolution of the density matrix after an initial time by
\begin{align}
\label{eq:decfree0}
\pdv{t} \rho_{0} &\approx \left(\mathcal{L}_0-P_0 \mathcal{L}_1 \mathcal{L}_0^{-1} \mathcal{L}_1\right) \rho_0,
\end{align}
where $P_0$ is the projection to the dissipation free subspace of $\mathcal{L}_0$. 

A substantial simplification occurs when the system described by $H_b(\lambda)$ is interacting and has the property that it thermalizes.
In this case, to describe local observables we approximate the  density matrix by a thermal state, $\rho^b \sim \exp {-\beta H_b(\lambda)}$.
This is justified if the thermalization time is short compared to the time-scale induced by 
scattering from photon fluctuations. This is the case for the relevant observables for $L\to \infty $ as the latter  time scale is proportional to $L/g^2$. 

\begin{figure}[t]
 	\centering
 	\includegraphics[width=0.46\textwidth]{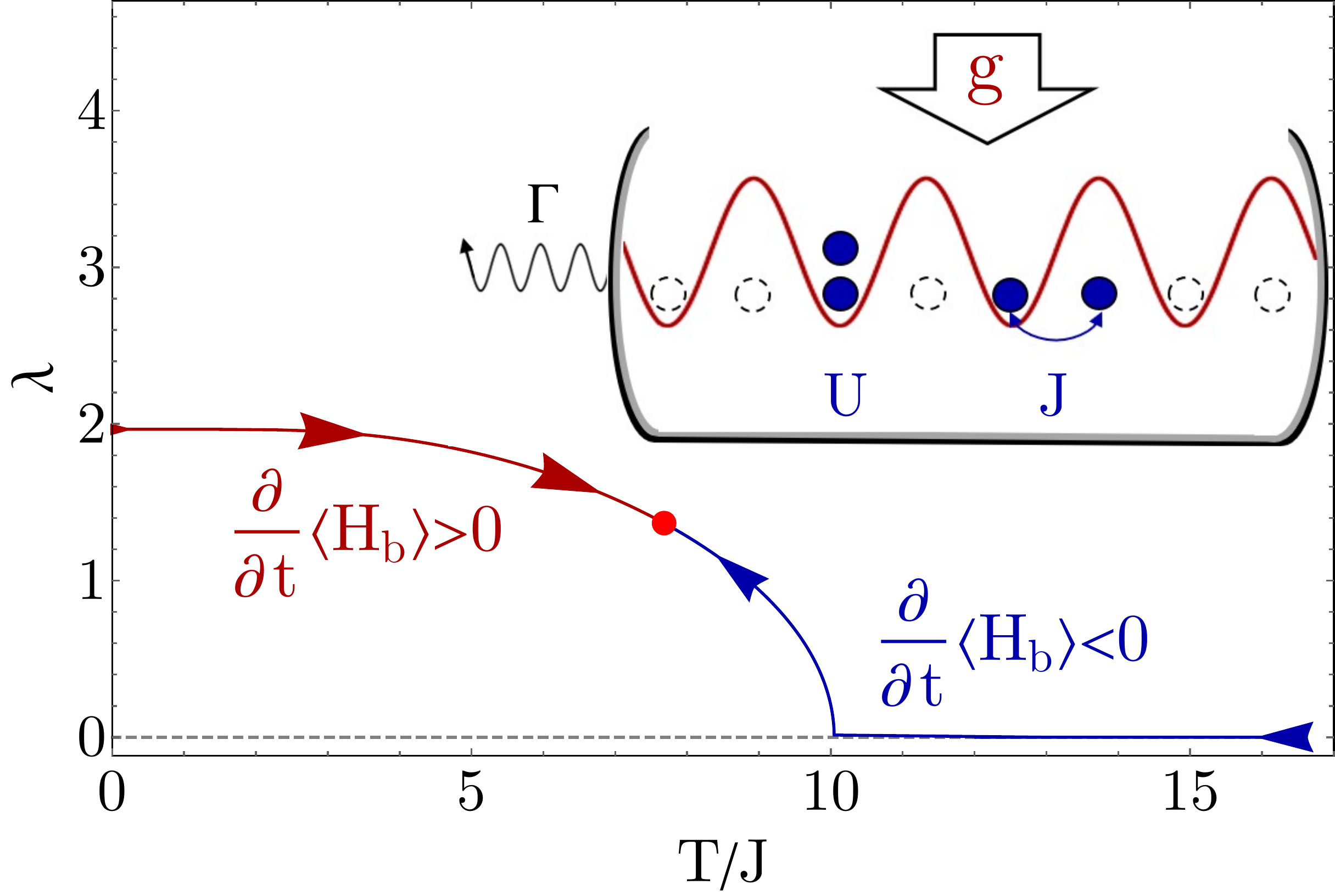}
 	\caption{Mean-field order parameter $\lambda$ as function of temperature computed for $\hbar g=3J$, $\hbar \Gamma=J$, $U=2J$, $L=10$. The temperature of the stationary state  is obtained by searching 
 	for the point (red dot) with $\frac{\partial}{\partial t}\langle H_b \rangle=0$. Here fluctuations beyond mean-field determine  $\frac{\partial}{\partial t}\langle H_b \rangle$. The inset shows a schematic picture of the 
 	 model: a Bose-Hubbard chain coupled to a cavity with loss rate $\Gamma$.}
 	\label{fig:order-param}
 \end{figure}

Further advantages of this Ansatz is that it directly cures the problem of positive definitness \footnote{Due to the applied perturbative expansion, the obtained density matrix does not need to be a physical density matrix, since the condition of positive definiteness might not be fulfilled. 
  \cite{LiKoch2014}}.
  The only remaining parameter is the temperature $T$ which can be computed from  
$\langle {H}_b \rangle$. Using Eq. \eqref{eq:decfree0} we obtain
 \begin{eqnarray}
 \label{rate_eq}
 \left\langle  \frac{\partial {H}_b}{\partial t} \right\rangle&=&
 \frac{2\hbar g^2 }{L}\! \int \!d\omega (1+n_B(\hbar\omega))\, \omega \, \text{Im}\chi^R(\omega) \delta_\Gamma(\omega+\delta)\nonumber, \\
 \delta_\Gamma(\omega)&=&\frac{\Gamma/(2 \pi)}{\omega^2+(\Gamma/2)^2}
 \end{eqnarray} 
 where $\chi^R(\omega)$ is the retarded correlation function of the operator $O$ calculated for a thermal state of the Hamilitonian $H_b$ (see Supplementary Material, section~\ref{app:heating}), while $\delta_\Gamma(\omega+\delta)$ describes the spectral function of the cavity mode broadened by the dissipation strength, $\Gamma$. 
 Cooling in Eq.~\eqref{rate_eq} arises from $\omega<0$, where the integrand is always negative, while heating comes from the positive integrand for $\omega>0$. 
Low-$T$ states heat up, see Fig.~\ref{fig:order-param}, while high-$T$ states are cooled towards a stable fixed point.
The time scale needed to reach the stationary state is proportional to the system size $L$ as the fluctuation driving the heating (or cooling) scale with $1/L$ (Eq.~\eqref{rate_eq}). 

Assuming the thermal state,  we simultaneously solve  $\left\langle  \frac{\partial}{\partial t} {H}_b\right\rangle=0$ and the mean-field equations~\eqref{eq:MF2} to obtain both  $T$ and the odd-even imbalance, $\Delta$, or, equivalently, the cavity field $\lambda$ of the steady state. 
Thus, we have reduced the problem of solving the Liouville equation, to the computation of the dynamical susceptibility $\chi^R(\omega)$ and the expectation value of energy for a thermal state of the Hamiltonian $H_\text{b}(\lambda)$.


 In the following we apply this  method to interacting bosonic atoms in one dimension 
  coupled to a single cavity mode transversely pumped with a standing-wave laser beam \cite{RitschEsslinger2013, MaschlerRitsch2008, NagyDomokos2008}, see inset of Fig.~\ref{fig:order-param},
\begin{align}
\label{eq:ham}
H &= H^0_\text{c}+H^0_\text{b}+H_{\text{bc}}\\
   H^0_b &=-J\sum_{j=1}^{L-1}(b_j^{\dagger}b_{j+1}+b_{j+1}^{\dagger}b_j)+\frac{U}{2}\sum_{j=1}^{L}n_j (n_j-1), \nonumber\\
H^0_c &=\hbar \delta\, a^{\dagger}a,~ H_{\text{bc}} =-\frac{\hbar g}{\sqrt{L}}(a+a^{\dagger})O,~
O=\sum_{j=1}^{L}(-1)^j n_j. \nonumber
 \end{align}
with hopping amplitude $J$ and a repulsive interaction of strength $U$. $\hbar\delta$ is the detuning between the cavity mode and the pump beam frequency obtained in the rotating frame.
Here we model a system~\cite{MaschlerRitsch2008} where the cavity field 
in combination with the pump laser field create a staggered potential described by the operator $O$. The strength of the cavity-boson coupling, $\hbar g/\sqrt{L}$, can thus be controlled by the pump laser. 

In practice, we use either exact diagonalization or, in some limits, analytical calculations to determine both $\Delta$ and $\left\langle  \frac{\partial}{\partial t} {H}_b\right\rangle$. Here it is important to note that the underlying mean field approximation should become exact in the thermodynamic limit, but we evaluate the equations with exact diagonalization of $H_b$ for rather small systems. This induces some finite-size errors. Luckily, those errors turn out to be very small as they are strongly suppressed due to the broadening induced by $\Gamma$ and the relatively high effective temperatures  which we obtain for most parameters, see below.

\begin{figure}[t]
	\centering
	\includegraphics[width=0.5\textwidth]{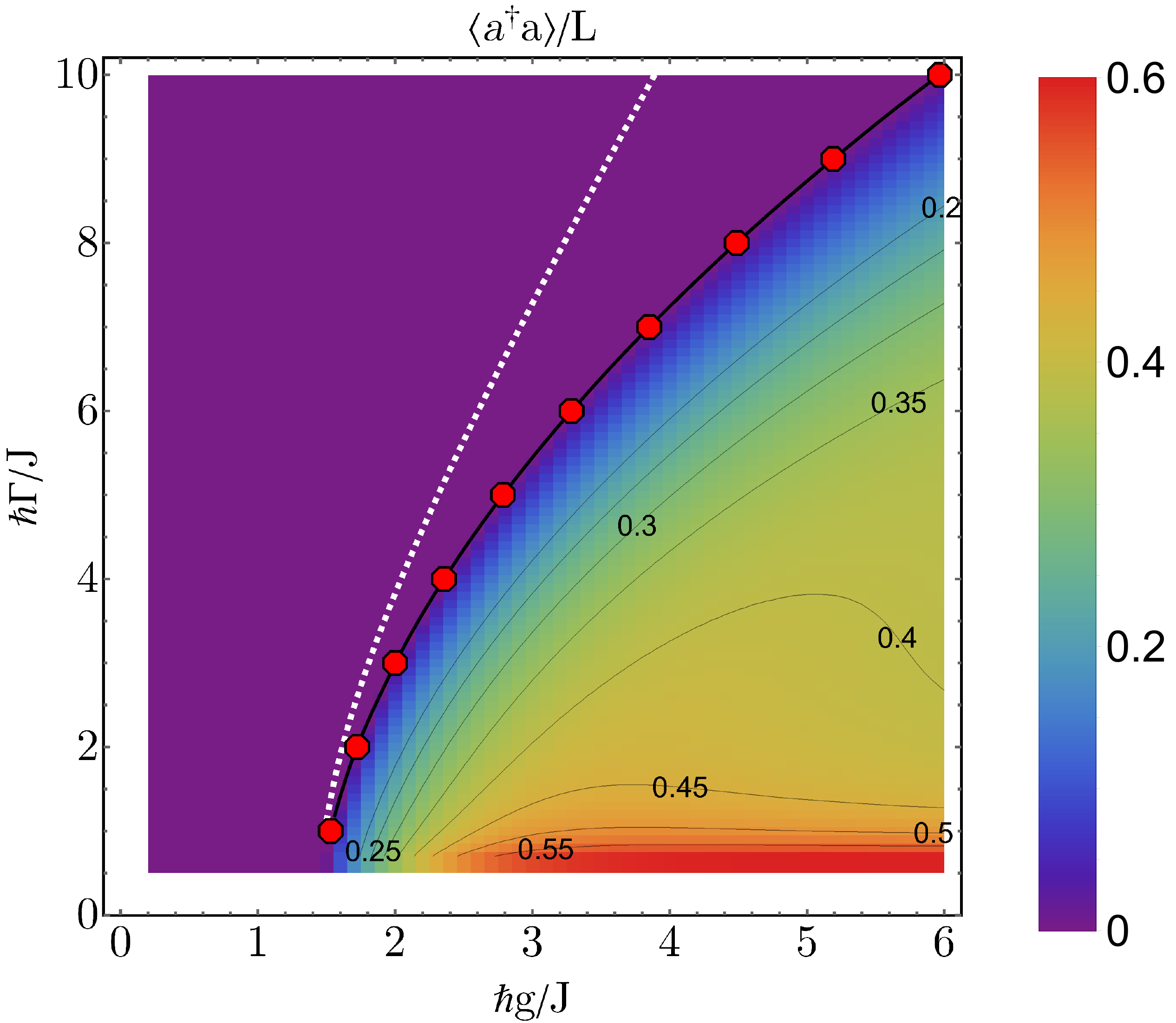}
	\caption{Phase diagram as function of $\Gamma$ and $g$ for $\hbar \delta=2 J$, $U=2 J$, $n=1/2$. The color and contour lines encode $\langle a^\dagger a\rangle/L$ which vanishes in the disordered phase for $L\to \infty$ and is finite in the ordered phase. Remarkably, $\langle a^\dagger a\rangle/L$ reaches the universal value $3/8=0.375$ deep in the ordered phase due to a heating effects, see Eq.~\eqref{universal_deep}. 
	}
	\label{fig:pd-lambda}
\end{figure}

In Fig.~\ref{fig:pd-lambda} the phase diagram is shown.
Close to the transition $g>g_c$ to the self-organized phase the photon number  $\frac{\langle a^\dagger a\rangle}{L}$ grows proportional to $g-g_c$ for $g>g_c$ and
saturates for $g\to \infty$. Analytically, we find that deep in the superradiant phase  the effects of heating compensates exactly the effect of a growing $g$ in the photon number (see Supplementary Material, section~\ref{app:heating})
  \begin{equation}\frac{\langle a^\dagger a \rangle}{L} \approx \frac{1}{2} n(1+n) \qquad \text{for } g \to \infty. \label{universal_deep}
 \end{equation}
In Fig.~\ref{fig:pd-lambda} we consider $n=1/2$ and therefore obtain $3/8$. In contrast, zero-temperature mean-field theory predicts $\frac{\langle a^\dagger a\rangle}{L} \approx \frac{g^2 n^2}{\delta^2+(\Gamma/2)^2}$ growing quadratically with $g$.
 
The increase of the critical $g_c$ with $\Gamma$, see Fig.~\ref{fig:pd-lambda}, deviates from the zero temperature mean-field prediction (white dashed line). Analytically we obtain from a high-temperature expansion for large $\Gamma$ or $\delta$ (see Supplementary Material, section~\ref{app:heating})
\begin{align}\label{gcAna}
g_c \approx \frac{\delta^2+(\Gamma/2)^2}{\delta \sqrt{8 n (1+n)}} \qquad \text{for}\ \hbar\Gamma \gg J, U \ \text{or }\ \hbar\delta \gg J, U.
\end{align}
Within a zero-temperature mean-field approximation, in contrast,  the critical coupling  is predicted to grow proportional to $\sqrt{1+\frac{\Gamma^2}{4 \delta^2}}$ instead.

\begin{figure*}[t]
 	\centering
 	\includegraphics[width=\linewidth]{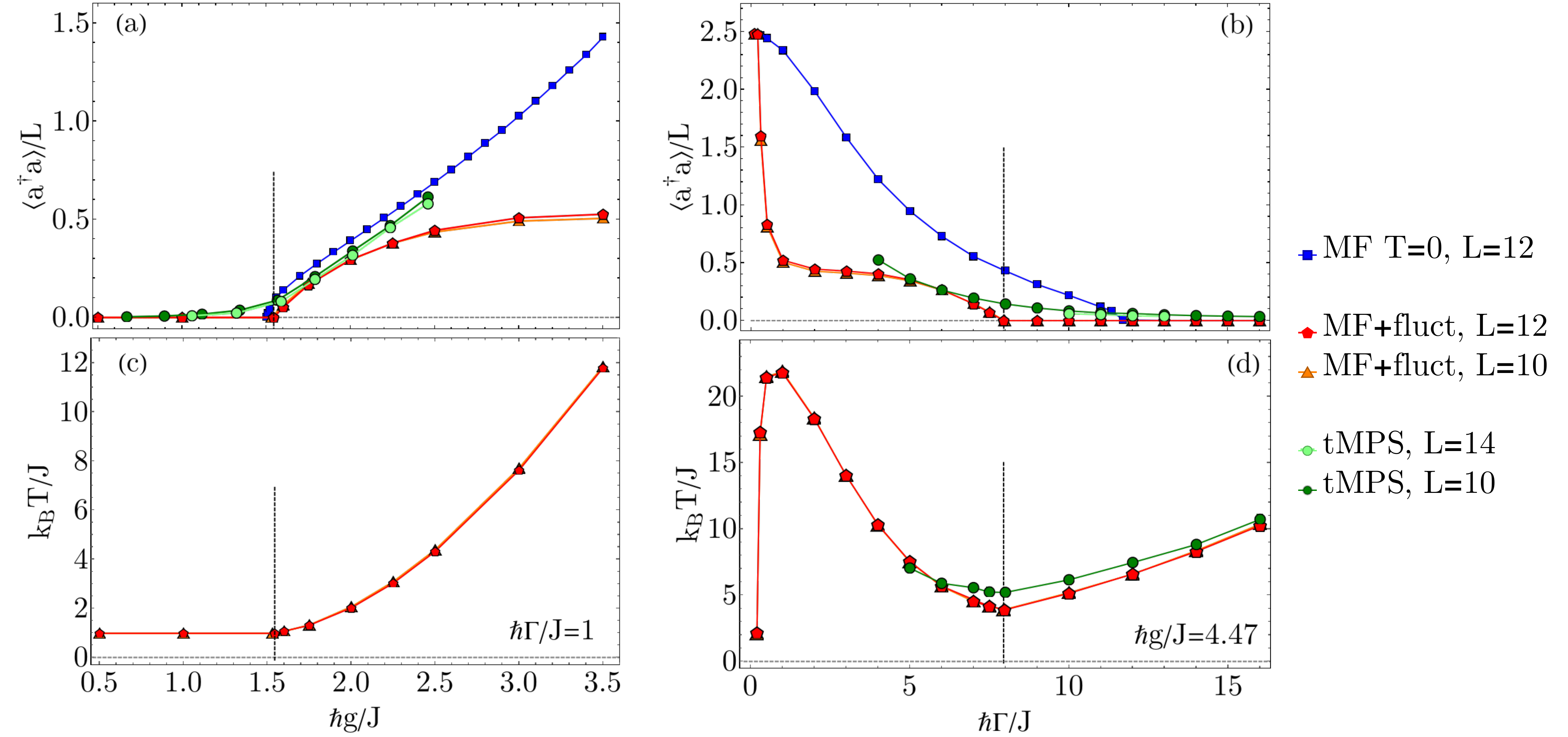}
 	\caption{$g$ and $\Gamma$  dependence of the photon number (panels (a) and (b)) and the effective temperature (panels (c) and (d)) for $U=2J$ and $\hbar \delta=2 J$. The mean-field results, Eq.~\eqref{eq:MF_cavity}, (red and orange for $L=12$ and $L=10$, respectively) are obtained using fluctuation effects to determine $T$, Eq.~\eqref{rate_eq}. 
 	The resulting photon numbers deviate strongly from the $T=0$ mean-field result (blue squares) due to the  heating effects shown in the two lower panels.
Numerically exact results obtained from $t$MPS 	(dark and light green circles) are fully consistent with the mean-field results where we only show parameters where finite-size effects are not too large. As expected, the sharp phase transition obtained within mean-field (vertical dashed lines) is smeared in the finite-$L$ $t$MPS calculations. }
 	\label{fig:all}
 \end{figure*}

In Fig.~\ref{fig:all} the number of photons and the effective atomic temperatures are shown for two cuts through the phase diagram. Upon increasing $g$ (Fig.~\ref{fig:all} (a), triangles and pentagons), a finite photon number $\langle a^\dagger a \rangle/L$ arises for  $g>g_c$ signaling a transition to the superradiant phase. Compared to the $T=0$ mean field approach (blue squares),
the rise of $\langle a^\dagger a \rangle$ is strongly suppressed for large $g$. The reason is the strong increase of $T$ in the ordered phase shown in Fig.~\ref{fig:all} (c). Note that $T$ is finite at $g=g_c$.
Fluctuations beyond mean-field induce even for $g<g_c$ a temperature with $k_BT/J\approx 1$. 

The strong influence of the temperature is also evident in Fig.~\ref{fig:all} (b) and (d) where the transition is studied as a function of $\Gamma$. For these parameters, the transition to the self-organized states below $\hbar \Gamma/J$ is considerably shifted from approximately $\hbar \Gamma/J\approx 12$ for the $T=0$ result to $\hbar \Gamma/J\approx 8$.

We have also performed numerical exact tMPS calculation for system sizes up to $L=14$ \cite{HalatiKollath2020,HalatiKollath2020b}.
In the regimes where we can reliably obtain steady-state properties, and in particular close to the transition thresholds, we find good agreement with the fluctuation-corrected mean field approach. Importantly, we can also extract an effective temperature from our tMPS result (see Supplementary Material, section~\ref{app:tmps}) which reproduces in Fig.~\ref{fig:all} (d) the characteristic minimum
of $T(\Gamma)$.


In important limits it is possible to calculate the steady-state temperature analytically (see Supplementary Material, section~\ref{app:heating} for details) and it is given by
 \begin{align}\label{Tasym}
k_B T \approx \left\{ 
 \begin{array}{ll} \frac{\hbar \delta}{\ln(1+\Gamma_0/\Gamma)} \qquad &\text{for}\ \hbar\Gamma \ll \hbar\delta, J, U\\
 \frac{\hbar (\Gamma/2)^2+\hbar(2 \lambda g-\delta)^2}{4\delta}&
 \text{for}\ \hbar g\lambda \gg J,U \text{ or } \hbar\delta \gg J,U\\
\frac{\hbar\kappa^2}{\delta} &
 \text{for}\, 0<\hbar\delta \ll \hbar\Gamma, J, U
 \end{array}\right.
 \end{align}
with \begin{align}
\Gamma_0&=\frac{2 \pi \delta \,\text{Im}\chi^R(\delta)}{\int_0^\infty\! d\omega\, \frac{\omega}{(\omega+\delta)^2} \text{Im}\chi^R(\omega)}\nonumber \\
\kappa^2&=\frac{\int \frac{\omega d\omega}{\omega^2+(\Gamma/2)^2} \text{Im} \chi^R(\omega)}{\int \frac{\omega d\omega}{(\omega^2+(\Gamma/2)^2)^2} \text{Im} \chi^R(\omega)}.
\end{align} 

These analytical expression can explain the evolution of $T$ shown in Figs.~\ref{fig:all} (c) and (d). The strong rise of $T$ in the ordered phase, 
Fig.~\ref{fig:all} (c),  is mainly driven by the increase of $g$, second line of Eq.~\eqref{Tasym}. Upon reducing $\Gamma$, Fig.~\ref{fig:all} (d), $T$ first drops as $T \sim \Gamma^2$ in this regime. Then it increases again because $\lambda$ grows rapidly in the ordered phase.  
In the limits of large detuning, $\delta$, and large dissipation strengths, $\Gamma$, we find similar scalings as in previous semiclassical approaches \cite{AsbothVukics2005, SchuetzMorigi2014, PiazzaStrack2014b}.
Finally, for $\hbar \Gamma\lesssim J$ and $\hbar \delta=U$,  a strong drop of the temperature (associated with a strong increase of photon number) can be seen.
 This is due to a special cooling mechanism only active for $\hbar \delta \approx U$: a doubly occupied state decays resonantly via the emission of a photon (see Supplementary Material, section~\ref{app:heating}).

To conclude, we have shown that in the thermodynamic limit a unique steady-state solution of interacting 
 bosons in a cavity can  be obtained only when fluctuation effects beyond mean field are taken into account.  Our method can be used for a large class of interacting many-particle systems coupled to a cavity model. 

 In case that the time scale for equilibrium $\tau_{eq}$ is shorter than the time scale for heating or cooling by fluctuations of the cavity field the state of the particles can be described by an effective temperature. 
In practice this situation is realized in most experiments.
 
 We have shown that in a large regime of parameters this temperature is very large in the steady state.
Even when tuning $\delta$ to a value optimized for cooling, Eqn.~\eqref{Tasym} predicts for 
 an arbitrary many-particle system with $\text{Im} \chi(\omega) \sim \omega^\alpha$, $\alpha\ge 1$, that the lowest possible temperature is proportional to  $\Gamma^{1/(1+\alpha)}$ and thus relatively large (prefactors are given in the Supplementary Material, section~\ref{app:heating}).
  Due to high-temperatures also an unexpected high degree of universality is obtained deep in the superradiant phase, leading, e.g., to an universal photon number, Eq.~\eqref{universal_deep}. We
expect that many of our results remain unaffected if, e.g., instead of one-dimensional bosons higher-dimensional realizations of our model \cite{KlinderHemmerich2015b, LandigEsslinger2016} are considered.

Our analysis has focused on the steady state obtained for $t \to \infty$ but  can easily be generalized to 
compute the time evolution \cite{Lange2018} on time-scales large compared to $\tau_{eq}$ and $1/\Gamma$ by computing the time evolution of the effective temperature. This is especially important as we expect that the experimental systems \cite{KlinderHemmerich2015b, LandigEsslinger2016} do not always reach the steady-state limit. 

Our approach is based on the computation of two equilibrium quantities: the thermal expectation value $\langle O \rangle$ and the linear-response susceptibility $\chi^R(\omega)$. Therefore it can directly be combined with a wide range of analytical and numerical approaches developed for closed quantum systems in equilibrium as, for example, Monte Carlo techniques, which can also be applied in higher dimensions.

 \acknowledgements
 We acknowledge useful discussions with Z. Lenar\v{c}i\v{c} and S. Diehl and funding from the German Research Foundation (DFG) under project number 277146847 - CRC 1238 (C04,C05), project number 277625399 - TRR 185 (B3) and under Germany’s Excellence Strategy – Cluster of Excellence Matter and Light for Quantum Computing (ML4Q) EXC 2004/1 – 390534769 and the European Research Council (ERC) under the Horizon 2020 research and innovation programme, grant agreement No. 648166 (Phonton). Furthermore, A.B. thanks the BCGS (Bonn-Cologne Graduate School of Physics and Astronomy) and the DAAD (German Academic Exchange Service) for support.

\appendix
\section*{Supplementary material}

 \section{Stationary states and analytical limits} \label{app:heating}
 To evaluate the change of energy, we evaluate the imaginary part of the retarded correlation function
 defined by $\chi^R(\omega)=-i \int e^{i \omega t}  \theta(t) \langle [Q(t),Q(0)] \rangle$ 
  using the Lehmann representation 
  \begin{align}
 \label{eq:lehmann}
 \text{Im}\chi^R(\omega)=&\sum_{n,m}   |\langle m| O |n\rangle|^2 
 \frac{e^{-\beta E_m}-e^{-\beta E_n}}{Z} \nonumber \\ 
 &\qquad \pi \delta\!\left(\omega-\frac{E_n-E_m}{\hbar}\right) 
 \end{align} 
 where $|n\rangle$ are eigenstates of $H_b$ with $H_b|n\rangle=E^b_n |n\rangle$ and $E_{nm}=E^b_n-E^b_m$.
 Note that $\text{Im}\chi^R(\omega)$ is an odd function, $\text{Im}\chi^R(-\omega)=-\text{Im}\chi^R(\omega)$.

 In several limits it is possible to compute the temperature of the steady state from Eq.~\eqref{rate_eq}  analytically. 
 In the analysis given below, we always assume that the density of bosons per site, $n$, is of order $1$. \\
 (1) $\Gamma \to 0$:  
 For small $\Gamma$, the cooling rate, i.e., the $\omega <0$ contribution of the integral in Eq.~\eqref{rate_eq}, is dominated by the value of $\omega = -\delta$, since the function $\delta_\Gamma(\omega+\delta)$ is sharply peaked at $\omega=-\delta$. This gives rise to a cooling rate proportional to $\left[1+n_B(-\hbar\delta)\right] \delta \, \text{Im}\chi^R(\delta)\approx - e^{-\hbar \beta \delta} \delta \, \text{Im}\chi^R(\delta)$. For the last approximation, we anticipated that $T$ becomes smaller and smaller in the limit $\Gamma \to 0$, see below. Heating arises from the $\omega\ge 0$ contributions. Using that $1+n_B(\hbar\omega)\approx 1$ for $\omega \ge 0$ at low $T$ and $\delta_\Gamma(\omega)\approx \frac{\Gamma/(2 \pi)}{(\omega+\delta)^2}$ for $\omega\ge 0$, we obtain a heating rate proportional to
 $\frac{\Gamma}{2 \pi} \int_0^\infty\! d\omega\, \frac{\omega}{(\omega+\delta)^2} \text{Im}\chi^R(\omega)$. Balancing heating and cooling, we obtain for the steady state the temperature
 \begin{align}\label{smallGamma}
 k_B T\approx\frac{\hbar \delta}{\ln(1+\Gamma_0(\delta)/\Gamma)} \qquad \text{for}\ \Gamma \ll J, U, \hbar \delta
 \end{align}
 with \begin{align}
 \Gamma_0(\delta)=\frac{2  \delta \,\text{Im}\chi^R(\delta)}{\int_0^\infty\! \frac{d\omega}{\pi}\, \frac{\omega}{(\omega+\delta)^2} \text{Im}\chi^R(\omega)}.\end{align}
  Note that the temperature decreases only logarithmically with $\Gamma$ for $\Gamma \to 0$ and the formula is only valid
 if  $\text{Im}\chi^R(\delta)\neq 0$. 

For a cavity of a given finesse or, equivalently, for a fixed value of $\Gamma$, one can use Eq.~\eqref{smallGamma} also to find the minimal temperature which can be achieved by a fine-tuning of the frequency of the pumping laser and thus of $\delta$. As $\Gamma_0/\delta$ vanishes for $\delta \to 0$, 
Eq.~\eqref{smallGamma} predicts that $T$ grows both for $\delta \to 0$ and for large $\delta$ thus predicting a minimal temperature as function of $\delta$. Here we assume that $\text{Im}\chi^R(\delta) \approx c_\chi \,\delta^\alpha$ for small $\delta$ and $T$, where $c_\chi$ is a constant. Furthermore, we define (motivated by the Kramers-Kronig relation) an effective cutoff frequency $\omega_c$ by $c_\chi\, \omega_c^\alpha=\chi^R(0)/2= \int_0^\infty \frac{\text{Im}\chi^R(\omega)}{\omega} \frac{d \omega}{\pi}$. With this definition we can approximate $\Gamma_0\approx 2 \delta (\delta/\omega_c)^\alpha$. By minimizing Eq.~\eqref{smallGamma} we find
\begin{align}
k_B T_{\text{min}} = c_\alpha \hbar \omega_c \left(\frac{\Gamma}{\omega_c}\right)^{1/(1+\alpha)}\quad \text{for } \Gamma\to 0.  \label{tmin}
\end{align}
with a numerical prefactor $c_\alpha = \min_x \frac{(x/2)^{1/(1+\alpha)}}{\ln[1+x]}$ which evaluates to $c_1\approx 0.88$ while $c_{\alpha \to \infty} \approx e/\alpha$. The optimal value of the detuning frequency is given by $\delta \approx \alpha c'_\alpha k_B T_{\text{min}}/\hbar$ with $c'_1\approx 2.1$ and $c'_{\alpha \to \infty} \approx 1$.

For ohmic friction, $\alpha=1$, realized, for example, in metals the minimally achievable temperature is proportional to $\sqrt{\Gamma}$ and thus remarkably large. For systems with a smaller density of states and thus a larger value of $\alpha$, $T_{\text{min}}$ will be even larger.
In many systems, e.g., a Bose Einstein condensates coupling via staggered potential to the cavity,
 $\text{Im} \chi^R(\omega)$ has a gap $E_g$ for $T\to 0$ which implies that $\alpha \to \infty$. In this can either chose $\hbar \delta$ to be of the order of the gap and use Eq.~\eqref{smallGamma}
with  $\hbar \delta \approx E_g$ or chose a smaller value of $\delta$ operating in a limit where $\text{Im}\chi^R(\delta)\sim e^{-\beta E_g}$. In both cases one obtains $k_B T_{\text{min}} \sim E_g$ with only logarithmic corrections. We therefore conclude that even in high-finesse cavity it is difficult to reach low temperatures in the steady state.

 (2) $\hbar\delta\gg J, U, \hbar g \lambda$ or $\hbar\Gamma, \frac{\hbar\Gamma^2}{\delta} \gg J, U, \hbar g \lambda$: In the limit of large detuning $\delta$ for an arbitrary ratio of $\Gamma/\delta$ or in the limit of sufficiently large $\Gamma$, we can use that $\text{Im}\chi^R(\omega)$ decays rapidly for frequencies larger than $J/\hbar, U/\hbar, g \lambda$. Therefore, one can approximate  $\delta_\Gamma(\omega+\delta)\approx
 \frac{\Gamma/(2 \pi)}{\delta^2+(\Gamma/2)^2}- \omega \frac{\Gamma \delta/\pi}{\left[\delta^2+(\Gamma/2)^2\right]^2}$. Furthermore for $k_B T \gg J, U, \hbar g \lambda$,
 we can use the high-temperature expansion, $\left[1+n_B(\hbar\omega)\right]\approx \frac{k_B T}{\hbar \omega}+\frac{1}{2}$. Using that $\omega \text{Im}\chi^R(\omega)$ is an even function of $\omega$, only the even terms of the product $\left[1+n_B(\hbar\omega)\right] \delta_\Gamma(\omega+\delta)$ survive which are given, to leading order, by the constant $\frac{1}{2}\frac{\Gamma/(2 \pi)}{\delta^2+(\Gamma/2)^2}-\frac{k_B T}{\hbar} \frac{\Gamma \delta/\pi}{\left[\delta^2+(\Gamma/2)^2\right]^2}$. This sum vanishes for the steady state which therefore is obtained for the temperature
  \begin{align}\label{largeDelta}
 k_B T\approx &\frac{\hbar\delta^2+\hbar(\Gamma/2)^2}{4 \delta} \\ 
 &\qquad \text{for}\ \hbar\delta \gg J, U, \hbar g \lambda \ \text{or}\ \hbar\Gamma, \frac{\hbar\Gamma^2}{\delta} \gg J, U, \hbar g \lambda. \nonumber
 \end{align}
 Remarkably, the temperature is completely independent of the parameters for the microscopic Hamiltonian of the atoms. 
Similar formulas for  effective temperatures have been derived within semiclassical approaches \cite{AsbothVukics2005, SchuetzMorigi2014, PiazzaStrack2014b} for {\em noninteracting} atoms coupled to a lossy cavity.  The absolute heating and cooling rates do depend on the microscopic parameters via the integral $\int  d\omega\, \omega \text{Im}\chi^R(\omega)$ but this factor cancels when one determines the steady state. The formula~\eqref{largeDelta} is  valid  for a broad range of other interacting Dicke models with large detuning $\delta$. Note that according to Eq.~\eqref{largeDelta}, the large temperature $k_B T\approx \hbar \delta/4$ is obtained in the small $\Gamma$ limit. This is not in contradiction to Eq~\eqref{smallGamma} as $\text{Im}\chi^R(\delta)$ and therefore $\Gamma_0$ vanishes exponentially for large $\delta$ which leaves the validity regime of Eq.~\eqref{smallGamma}.

 \begin{figure}[b]
 	\centering
 	\includegraphics[width=0.9\linewidth]{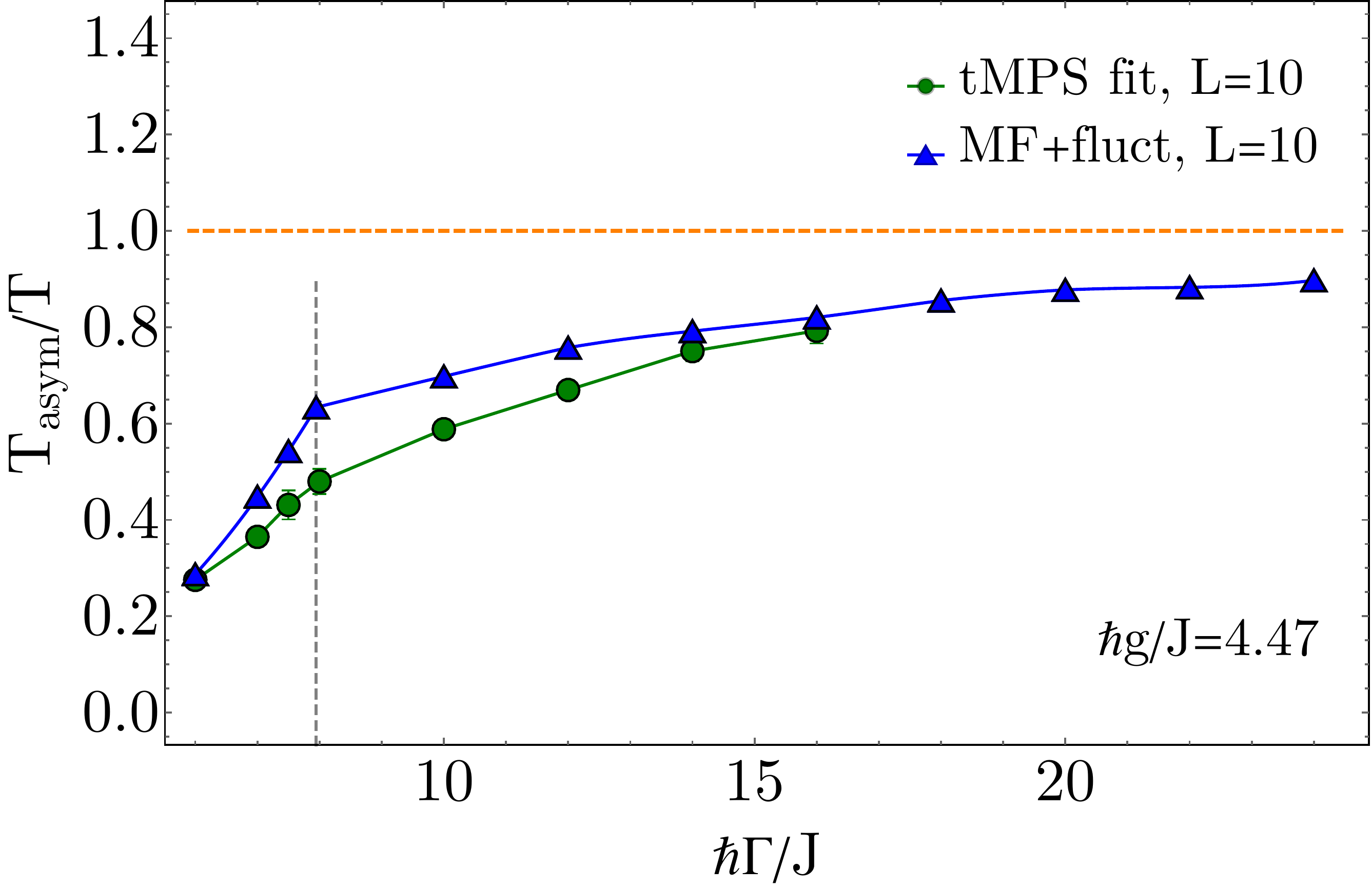}
 	\caption{Plot for the ratio of the asymptotic value for $T$ given by the Eq.~(\ref{largeDelta}) and $T$ obtained from the fluctuation-corrected mean field (blue triangles) and tMPS calculation (dark green circles). The grey dashed line signifies the phase transition ($U=2J$, $\hbar \delta = 2J$).\label{fig:asymptotic2}}.
 \end{figure}
  In Fig.~\ref{fig:asymptotic2} we plot the ratio of the asymptotic value of $T$ from Eq.~\eqref{largeDelta} to the  temperature obtained from our fluctuation-corrected mean-field theory. As expected the ratio approaches $1$ for large $\Gamma$.

 As most experiments are performed in the large $\delta$, large $\Gamma$ limit, it is also useful to estimate how long it takes to reach the large temperatures predicted by Eq.~\eqref{largeDelta}. For this we assume that $T$ is initially much smaller than the steady state temperature. In this case, one can approximate  $\delta_\Gamma(\omega+\delta)\approx
 \frac{\Gamma/(2 \pi)}{\delta^2+(\Gamma/2)^2}$ (ignoring the next order term as it is much smaller for low $T$). Furthermore, as $n_B(\hbar\omega)+n_B(-\hbar\omega)=-1$, we obtain $\int (1+n_B(\hbar \omega)) \omega \text{Im}\chi^R(\omega)=\frac{1}{2} \int \omega  \text{Im}\chi^R(\omega)$. The latter term is computed using the 
 sum rule $\int  d\omega\, \omega \,\text{Im}\chi^R(\omega) = \pi \langle [[O,H_b],O] \rangle = -4 \pi \langle H_{\text{kin}} \rangle$ where  $H_{\text{kin}}$
 is the kinetic energy of the bosons proportional to $J$, the only term in $H_b$ not commuting with $O$. The factor $4=2^2$ arises because a single hopping term changes the staggered potential $O$ by $2$. The initial heating rate at low $T$ is therefore given by
 \begin{align}
 \left\langle \frac{\partial H_b}{\partial t} \right\rangle &\approx   \frac{2  g^2}{L}  \frac{\Gamma}{\delta^2+(\Gamma/2)^2} \langle -H_{\text{kin}}\rangle \\
 & \text{for } \hbar \delta,\hbar \Gamma \gg J,U, \hbar g \lambda\  \text{and}\  k_B T \ll \frac{\hbar \delta^2+\hbar(\Gamma/2)^2}{4 \delta}. \nonumber
 \end{align}
This result is valid independent of the dimension of the system assuming a staggered potential on a bipartite lattice. One can use the formula to estimate how long it takes to add an energy of order  $\Delta E=\langle -H_{\text{kin}}\rangle$ to the system. This happens after a time of order 
 $t \sim \frac{L}{g^2} \frac{\delta^2+(\Gamma/2)^2}{\Gamma}$ which is of order $\frac{L}{\Gamma}$ for $g$ close to the critical coupling,  $g\sim \delta$, assuming $\Gamma \lesssim \delta$.

(3) $0<\hbar\delta \ll \hbar\Gamma, U, J$: For $\delta \to 0$, less and less energy is removed from the system when a photon leaves the cavity. And we find that the temperature increases in this limit.
  Thus, we perform a simultaneous high-temperature expansion and Taylor expansion in $\delta$ in this limit.  As above, only the even part of $\left[1+n_B(\hbar\omega)\right] \delta_\Gamma(\omega)$ contributes to the heating rate~\eqref{rate_eq} which is proportional to 
 $\frac{1}{2 \left[\omega^2+(\Gamma/2)^2\right]}-\frac{2 k_B T \delta}{\hbar \left[\omega^2+(\Gamma/2)^2\right]^2}$.
 Therefore, the steady state temperature can be approximated by 
 \begin{eqnarray}
k_B T \approx \frac{\hbar\kappa^2}{4 \delta} \quad \text{for } 0<\hbar\delta \ll J,U, \hbar g \lambda
 \end{eqnarray}
 with $\kappa^2=\frac{\int \frac{\omega d\omega}{\omega^2+(\Gamma/2)^2} \text{Im} \chi^R(\omega)}{\int \frac{\omega d\omega}{\left[\omega^2+(\Gamma/2)^2\right]^2} \text{Im} \chi^R(\omega)}$ fully consistent with Eqs.~\eqref{largeDelta}
 and \eqref{largeLambda} which are valid for large $\Gamma$ and large $g \lambda$, respectively. Importantly, we find that the temperature always diverges in the limit $\delta \to 0$. Low temperatures can only be reached when both $\Gamma$ and $\delta$ are reduced simultaneously, see Eq.~\eqref{smallGamma}.

(4) $\hbar g \lambda \gg J, U$: Deep in the ordered self-organized phase, the staggered potential $\hbar g \lambda$ becomes much larger than 
 hopping and interactions. In this limit the staggered potential dominates the energy. A hopping process changes the energy by $2 \hbar g \lambda$ and therefore one finds (independent of $T$) that $\text{Im}\chi^R(\omega) \propto \delta(\omega-2 g \lambda)- \delta(\omega+2 g \lambda)$ (further peaks at $\hbar\omega=\pm m U$, $m \in \mathbb N$, are discussed below). The overall prefactor proportional to $J^2/(\lambda g)$ does not influence the steady state and is therefore omitted here. Thus, the change of energy is proportional to $\left[1+n_B(2 \hbar g \lambda)\right] \delta_\Gamma(2 g \lambda+\delta)+\left[1+n_B(-2\hbar g \lambda)\right] \delta_\Gamma(2 g \lambda-\delta)$.  This prefactor vanishes for
 \begin{align}\label{largeLambda}
 k_B T=\frac{2 \hbar g \lambda}{\ln\!\left[1+\frac{8 g \lambda \delta}{(\Gamma/2)^2+(2 g \lambda-\delta)^2}\right]}\approx &\frac{\hbar(\Gamma/2)^2+\hbar(2 g \lambda-\delta)^2}{4\delta}\nonumber\\
 &\text{for}\ g\lambda \gg J,U.
 \end{align}
 This equation is fully consistent with Eq.~\eqref{largeDelta}, which is recovered for $\delta \gg g \lambda$.

\begin{figure}[t]
 	\centering
 	\includegraphics[width=0.9\linewidth]{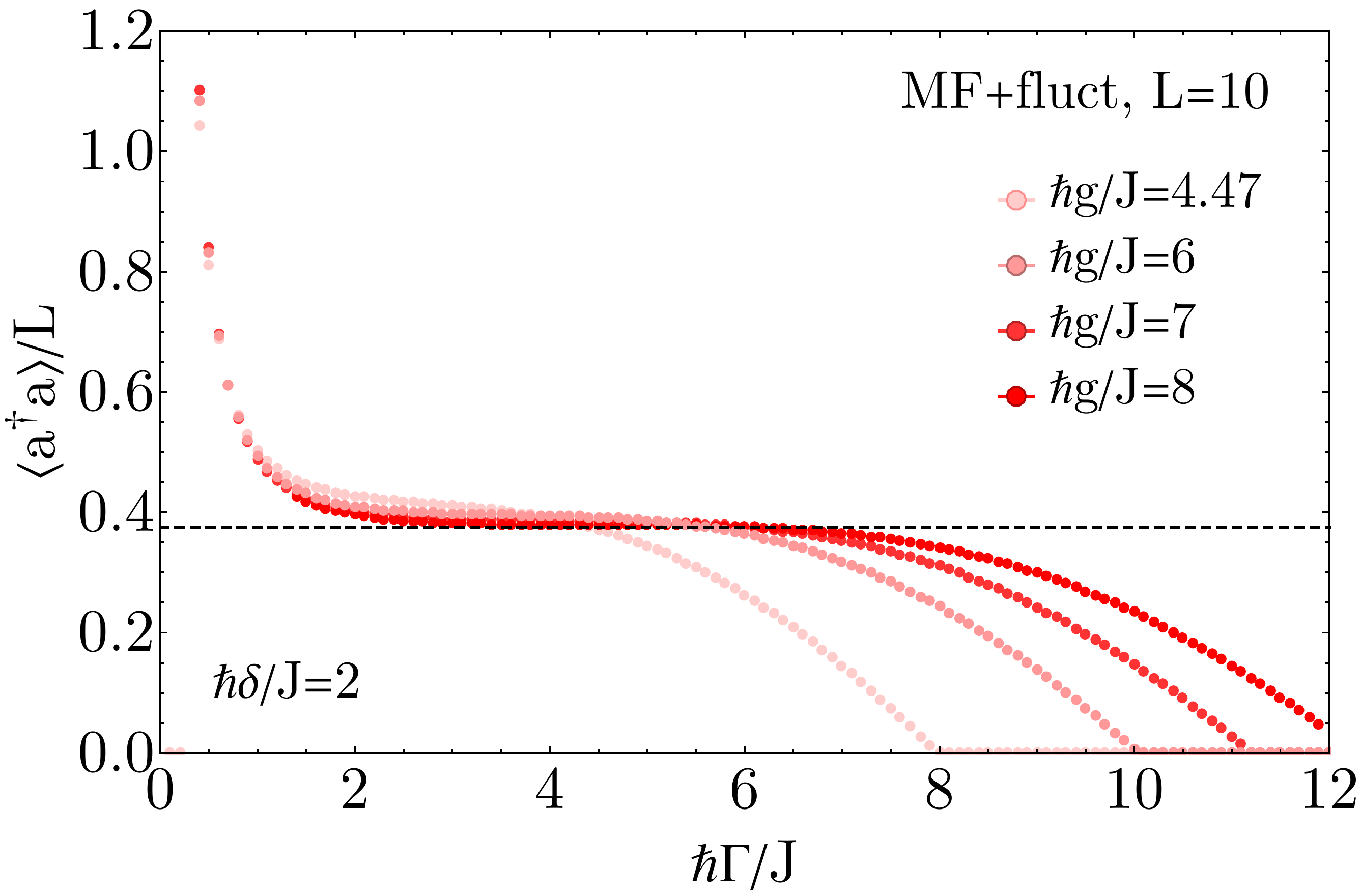}
 	\caption{Photon number in the large $g$ limit as function of $\Gamma$. For large $g$, the universal result  $\frac{\langle a^\dagger\rangle_c \langle a \rangle_c}{L}\approx \frac{1}{2} n (1+n)$ (dashed line) is obtained for a broad range of $\Gamma$.  \label{fig:universalPhoton}}
 \end{figure}

 Using a high-$T$, large $g \lambda$ expansion for the order parameter we obtain $\frac{\langle O \rangle}{L} \approx  \frac{\hbar \lambda g\, n(1+n)}{k_B T}$. Combining this with $k_B T\approx \frac{\hbar(2 g \lambda)^2}{4\delta}$ (assuming that $g \lambda \gg \Gamma,\delta$)
 from Eq.~\eqref{largeLambda}, we obtain from the mean field equations for $g \to \infty$
 \begin{align}
 k_B T&\approx \frac{2 \hbar g^2 n (1+n) \delta}{\delta^2+(\Gamma/2)^2}, \quad \lambda \approx \frac{\delta \sqrt{2 n (1+n)}}{\sqrt{\delta^2+(\Gamma/2)^2}}, \nonumber \\
  \frac{\langle a^\dagger a \rangle_c}{L} 
 &\approx \frac{\langle a^\dagger\rangle_c \langle a \rangle_c}{L}\approx \frac{1}{2} n (1+n).\label{nUniv}
 \end{align}
 Remarkably, $\langle a^\dagger a \rangle_c$ is completely independent of all microscopic parameters. In Fig.~\ref{fig:universalPhoton} we show how this universal value is approached when $g$ gets larger.

  (5) $\hbar g \lambda \gg J, U$ for $U\approx \hbar \delta$ and small $\Gamma$: A special situation arises deep in the ordered phase if the interaction strength $U$ matches the photon frequency $\delta$. In this case a resonant process provides an efficient extra cooling mechanism.
For   $\hbar g \lambda \gg J$, hopping processes are strongly suppressed by the large staggered field. As discussed above, 
there are pronounced peaks at $\text{Im}\chi^R(\omega)$ at  $\omega=\pm 2 g \lambda$ arising from processes where bosons hop between even and odd sites. There exist, however, also peaks arising from the hopping from even to even (or odd to odd) sites. Consider, for example an initial state with $n_i$ atoms on site $i$ and $n_{i+2}$ atoms on site $i+2$, which have the same local potential. By a second-order hopping process, an atom may hop from site $i$ to site $i+2$ resulting in states with $n_i-1$ and $n_{i+2}+1$ atoms on site $i$ and $i+2$, respectively. In this process, the interaction energy $\frac{U}{2} \sum_j n_j (n_j-1)$ changes by  $ m U$ with $m=1+n_{i+2}-n_i$. For $J^2/(\hbar g \lambda) \ll U$
 this leads to sharp peaks in $\text{Im}\chi^R(\omega)$  at frequencies $\hbar\omega=\pm m U$, $m \in \mathbb N$. Their weight is, however, suppressed by a factor of $\left[J/(\hbar g \lambda)\right]^2$ compared to the primary peaks at $\omega=\pm 2g \lambda$. Taking also the factor $\omega$ in Eq.~\eqref{rate_eq} into account, which also favors large energy transfers, one finds that the secondary peaks will become only important for large $g \lambda$ if there is a resonant coupling to the cavity mode at $\hbar \delta \approx m U$ and therefore {\em resonant cooling}.
 
For $J^2/(\hbar g \lambda)\ll U,\hbar\Gamma$, the resonant cooling rate for $m=1$, $\hbar \delta=U$ is therefore proportional to  $(1+n_B(-U) \left(\frac{J}{\hbar g \lambda}\right)^4  \frac{U}{\hbar\Gamma}$ to be compared to a heating rate proportional to $\left[1+n_B(\hbar g \lambda)\right] \left(\frac{J}{\hbar g \lambda}\right)^2 \hbar g \lambda \frac{\Gamma}{(\hbar g \lambda)^2}$. 
For $k_B T \gg \hbar g\lambda$ both rates are proportional to $1/(g \lambda)^4$ but cooling dominates compared to 
heating for $\hbar \Gamma \lesssim J$. 
In an intermediate regime where $1+n_B(\hbar g \lambda)\approx 1$ and $1+n_B(-U)\approx -\frac{k_BT}{U}$, one obtains
\begin{align}
k_B T \sim \hbar^3 \Gamma^2 \frac{g \lambda}{J^2}  \qquad \text{for}\ \left(\frac{U J^2}{\hbar g \lambda}\right)^{1/2}\!\!\!< \hbar\Gamma  <J, \hbar \delta=U
\end{align}
crossing over to Eq.~\eqref{smallGamma} for even smaller values of $\Gamma$.
The efficient resonant cooling is the reason why temperatures drop rapidly for small $\Gamma$ in Fig.~\ref{fig:all}.

\section{Details of the \MakeLowercase{t}MPS method for the coupled
photon-atom system}\label{app:tmps}

We compare our fluctuation-corrected mean field results with numerically exact results obtained with a matrix product state (MPS) method developed for the simulation of the time evolution of the dissipative master equation, Eqs.~(\ref{eq:lind}) and (\ref{eq:ham}), for the cavity-atoms coupled systems. The details regarding the implementation and benchmarking of the method are presented in Ref.~\cite{HalatiKollath2020b}. The method is based on the stochastic unravelling of the master equation with quantum trajectories and a variant of the quasi-exact time-dependent variational matrix product state (tMPS) employing the Trotter-Suzuki decomposition of the time evolution propagator 
and the dynamical deformation of the MPS structure using swap gates.

\begin{figure}[t]
\centering
\includegraphics[width=0.46\textwidth]{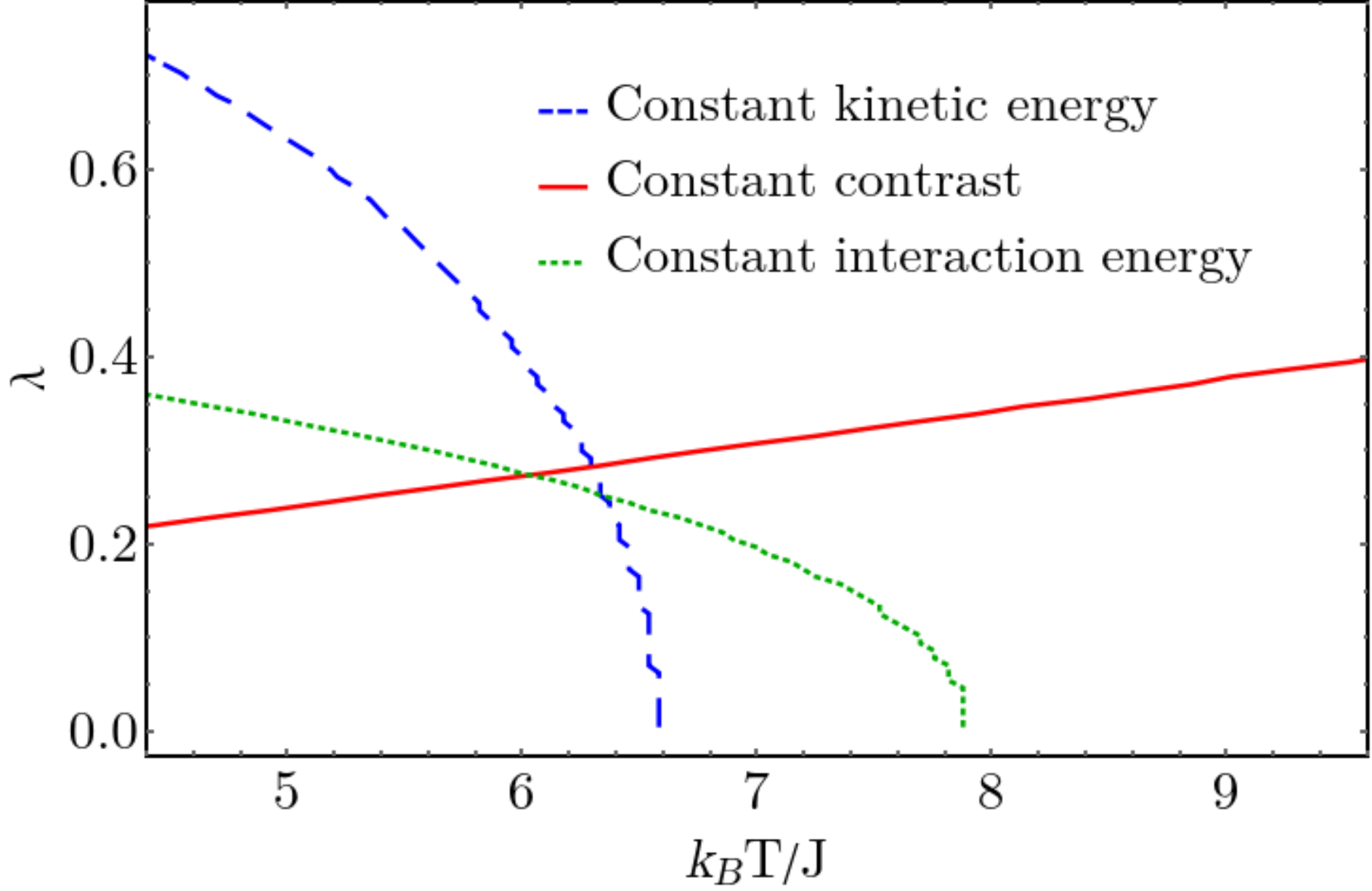}
\caption{Extraction of the effective temperature from the tMPS data for the parameters $L=10$, $N=5$, $\hbar g/J=4.47$, $\hbar\delta/J=2$, $U/J=2$, $\hbar\Gamma/J=10$, $tJ = 49.75 \hbar$. The lines in the $T$-$\lambda$ plane are for constant kinetic energy, $E_\text{kin}/J=-1.85$, contrast of the density-density correlations, $\frac{1}{L-2}\sum_j \left(\langle n_j n_{j+2}\rangle-\langle n_j n_{j+1}\rangle\right)=0.03$ and interaction energy, $E_\text{int}/J=2.75$.}
\label{fig:fits}
\end{figure}

The tMPS results presented in Fig.~\ref{fig:all} are taken at times of $tJ = 49.75 \hbar$. The convergence of our results is sufficient \cite{HalatiKollath2020b} for at least 500 quantum trajectories in the Monte Carlo sampling, the truncation error goal of $10^{-12}$ for $L=10$ and $10^{-9}$ for $L=14$, the time-step of $dtJ = 0.0125 \hbar$ or smaller, an adaptive cutoff of the local Hilbert space of the photon mode between
$N_\text{pho}= 35$  and $N_\text{pho}= 10$.

In order to the compare the values of the effective temperature obtained in the developed perturbation around the mean field approach we need to extract an effective temperature that can describe the tMPS results (see Fig.~\ref{fig:all}(d)). Thus, we try to find the parameters $T$ and $\lambda$, which determine the density matrix $\rho(T,\lambda)\sim\ket{\alpha(\lambda)}\bra{\alpha(\lambda)}e^{- H_{b}(\lambda)/k_B T}$, by requiring the thermal density matrix to approximately describe the tMPS results.
For this we employ the following procedure. We first compute the expectation values of three important observables of the atomic sector with tMPS, the kinetic energy, $E_\text{kin}=-J\sum_{j=1}^{L-1}\langle b_j^{\dagger}b_{j+1}+b_{j+1}^{\dagger}b_j\rangle$, the contrast of the density-density correlations, $\frac{1}{L-2}\sum_{j=1}^{L-2} \left(\langle n_j n_{j+2}\rangle-\langle n_j n_{j+1}\rangle\right)$ and interaction energy, $E_\text{int}=\frac{U}{2}\sum_{j=1}^{L}\langle n_j (n_j-1)\rangle$.

In the next step we compute the expectation values of the mentioned observables using $\rho(T,\lambda)$ and identify the points for which we obtain the same values as in tMPS. In Fig.~\ref{fig:fits}, we see that in the $T$-$\lambda$ plane for each observable we find a curve along which the expectation value agrees with tMPS. Thus by finding the intersection point of the three curves we obtain the values of $T$ and $\lambda$ for which $\rho(T,\lambda)$ approximates the tMPS state.
As we observe that these curves do not intersect each other in a unique point, but rather in three distinct ones, we define the area of the triangle as a measure the errors involved in the determination of the effective temperature, see Fig.~\ref{fig:fits}.
Note that the cavity field $\lambda$ used as a parameter in this procedure does not agree with the tMPS photon number, as in the full quantum evolution the cavity field is not restricted to a coherent state.
We note that for $\Gamma/J\lesssim 5$ this procedure gives unreliable results as the intersection points are far from each other.

\end{document}